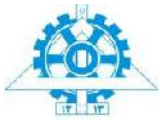
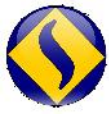
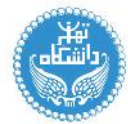

# Image Processing Code for Sharpening Photoelastic Fringe Patterns and Its Usage in Determination of Stress Intensity Factors in a Sample Contact Problem

S. Khaleghian[*], A. Emami and N. Soltani

*School of Mechanical Engineering, College of Engineering, University of Tehran, Tehran, Iran*

**Abstract**

This study presented a type of image processing code which is used for sharpening photoelastic fringe patterns of transparent materials in photoelastic experiences to determine the stress distribution. C-Sharp software was utilized for coding the algorithm of this image processing method. For evaluation of this code, the results of a photoelastic experience of a sample contact problem between a half-plane with an oblique edge crack and a tilted wedge using this image processing method was compared with the FEM results of the same problem in order to obtain the stress intensity factors (SIF) of the specimen. A good agreement between experimental results extracted from this method of image processing and computational results was observed.

## 1. Introduction

A practical technique to experimentally obtain stress distribution in transparent materials is photoelasticity. However, in the photoelastic experiments, the exact path of isochromatic fringes in images captured by professional cameras can be hardly detected by bare eyes, and this problem usually causes considerable errors in final image analysis. Moreover, the exact locations of fringe pixels are require in the fracture mechanics equations for calculation stress intensity factors of materials, and as theory of photoelasticity says an isochromatic fringe should be a line without width for which the stress-optic law is true.

One of the effective methods of eliminating the errors occurred in determination of pixels located on an isochromatic fringe is image processing. Up to now, many researches have been done to develop an image processing method for photoelastic fringe pattern analysis. In 1990, Wang Wei-Chung developed an algorithm on a digital photoelastic system for extracting stress intensity factor from the experimentally obtained photoelastic fringe pattern [1]. In the following year, A.C. Gillies worked on a low-level image-processing algorithm for photoelastic fringe pattern analysis [2]. Then in 1994, J. Carazo-Alvarez et al combined two automated systems of photoelastic analysis to identify the absolute value of the isochromatic parameter at a point with minor errors than previous techniques [3].

Later in 1998, S Yoneyama et al presented new methods of photoelastic fringe analysis from a single image [4-5]. Afterwards, many other researchers like T.Y. Chen et al [6] and C.W. Chang et al [7-8] utilized digital image processing to determine stress distribution in some practical mechanical problems.

This study presented a type of image processing code in which the exact locations of pixels in a fringe path can be detected. This code used a significant property of all the pixels that belong to isochromatic fringes. This communal property is their extremum light intensity which is the basis of this type of image processing algorithm.

The image processing code was utilized in photoelastic experimental analysis of a contact problem between a tilted asymmetric wedge and an infinite half plane containing an oblique edge crack to measure its stress intensity factors. Then, FEM analysis of the same problem was carry out for verification of the algorithm outcomes and evaluation of its effectiveness in reducing the errors of experimental results.

## 2. Image processing algorithm

The algorithm of this method of image processing was coded in C-Sharp software. At the first step of this algorithm, the image got divided into tiny segments. Then, the gradient of light intensity for each segment was calculated. Then, the segments with zero

---

[*]Corresponding author. Tel.: +98 912 7127609, Fax: +98 21 44078071

E-mail address: sm.khaleghian@yahoo.com

ICME 2011

gradient of light intensity were selected in the fringe list. If the gradient of segments in neighborhood of a listed segment was also zero, the listed segment would remain in the list; but if the adjacent segments of a segment in the list had non-zero gradient of light intensity, the segment would be removed from the list. This step of algorithm prevented the selection of a stain in the fringe list. After completion of the fringe list, all the adjoining segments in the list were connected by a line which showed the path of an isochromatic fringe which had the extremum light intensity.

### 3. Sample problem for verification of algorithm

In order to test the new algorithm and quantitatively assess its performance, this image processing technique was used in experimental determination of stress intensity factors (SIF) of a contact problem between a half-plane with an oblique edge crack and an asymmetric tilted wedge. The geometry of this contact problem was shown in Fig. 1 where "a" is crack length, $X_c$ is the distance between wedge tip and crack opening, h is width and L is length of plane.

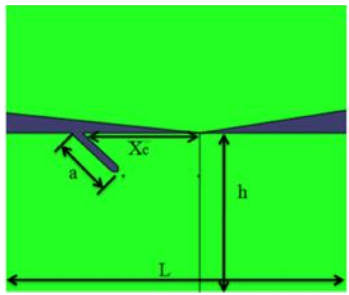

Fig. 1 Geometry of contact problem

In this investigation, the optimal required dimensions of the plate to mimic the infinite half plane were found to be L=240mm and h=60mm. The half-plane used in this study was made of Makrolon polycarbonate plates with 4mm thickness from Byer Company (Germany). The mechanical properties of this type of polycarbonate are shown in Table1. 45° Crack of 5mm long was machined using a 2mm blade into the plate, and the distance between wedge tip and crack opening was also 5mm.

Table 1 Table1- mechanical properties of specimens

| Material | E (MPa) | $\vartheta$ | Friction coefficient, $\mu$ |
|---|---|---|---|
| Makrolon Polycarbonate | 2400 | 0.38 | 0.4 |

The asymmetric tilted wedge used for contact was made of plexiglas, and the dimensions and angels of the asymmetric wedge used for contact are shown in Fig. 2. The asymmetric wedge was connected to a 200Kg.F load cell used for measuring the contact loading. The applied load on the load was measured 255KN.

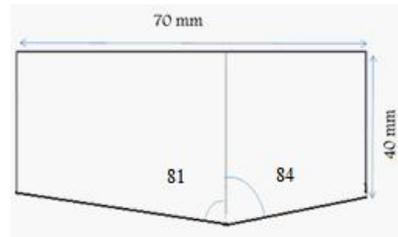

Fig. 2 Geometry of asymmetric wedge

### 3.1. Experimental analysis

Experimental data were acquired using a polariscope designed for a wavelength of 562 and the light source which was modified to generate the same wavelength. The image of photoelastic fringe pattern was captured by a 7-Mega pixel digital camera. Then, the image processing code was employed to the photoelastic fringe outputs. Captured photo of photoelastic fringes before image processing and after image processing is shown in Fig. 3 and Fig. 4 respectively.

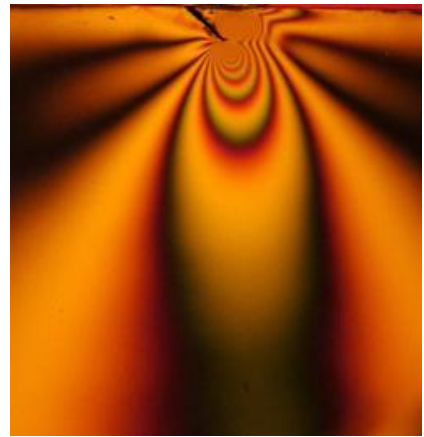

Fig. 3 Photo of photoelastic fringes before image processing

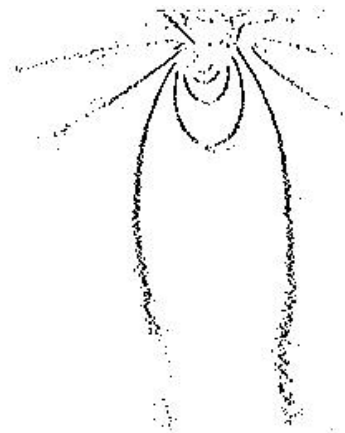

Fig. 4 Photo of photoelastic fringes after image processing

After image processing, the photo was loaded by a program written in MATLAB software which was able to detect the actual size of the model in loaded photo. This program was utilized to determine the actual polar coordinates of any selected pixel of the processed image. In the image loaded by this program, 20 pixels located on the path of crack tip fringes were selected and their actual polar coordinates were obtained which are shown in Table 2. The number of selected pixels was elected so that they gave the intensity factors with minimal error (less than 1%) [9].

Table 2. Position of selected pixels

| Point number | $r$ (m) | $\theta$ (rad) | $N$ (fringe order) |
|---|---|---|---|
| 1 | 0.024 | -1.07789 | 1 |
| 2 | 0.0211 | -1.12854 | 1 |
| 3 | 0.0196 | -1.15269 | 1 |
| 4 | 0.0176 | -1.19645 | 1 |
| 5 | 0.0167 | -1.21795 | 1 |
| 6 | 0.0151 | -1.22746 | 1 |
| 7 | 0.0121 | -1.29937 | 1 |
| 8 | 0.0097 | -1.35914 | 1 |
| 9 | 0.0085 | -1.40095 | 1 |
| 10 | 0.0068 | -1.43829 | 1 |
| 11 | 0.0058 | -1.21031 | 2 |
| 12 | 0.0067 | -1.15516 | 2 |
| 13 | 0.0077 | -1.09189 | 2 |
| 14 | 0.0085 | -1.03722 | 2 |
| 15 | 0.0091 | -0.97801 | 2 |
| 16 | 0.004 | -1.10367 | 3 |
| 17 | 0.0048 | -1.04021 | 3 |
| 18 | 0.0053 | -0.95138 | 3 |
| 19 | 0.0059 | -0.87171 | 3 |
| 20 | 0.0062 | -0.79798 | 3 |

Assuming a zero stress through thickness ($\sigma_t = 0$), stress filed around the crack tip can be defined as[10]:

$$\sigma_r = \frac{K_I}{4\sqrt{2\pi r}}(5\cos(\frac{\theta}{2}) - \cos(\frac{3\theta}{2})) + \frac{K_{II}}{4\sqrt{2\pi r}}(-5\sin(\frac{\theta}{2}) + 3\sin(\frac{3\theta}{2}))$$

$$\sigma_\theta = \frac{K_I}{4\sqrt{2\pi r}}(3\cos(\frac{\theta}{2}) + \cos(\frac{3\theta}{2})) + \frac{K_{II}}{4\sqrt{2\pi r}}(-3\sin(\frac{\theta}{2}) - 3\sin(\frac{3\theta}{2}))$$

$$\tau_{r\theta} = \frac{K_I}{4\sqrt{2\pi r}}(\sin(\frac{\theta}{2}) + \sin(\frac{3\theta}{2})) + \frac{K_{II}}{4\sqrt{2\pi r}}(\cos(\frac{\theta}{2}) + 3\cos(\frac{3\theta}{2}))$$

(1)

Where $K_I$ and $K_{II}$ are stress intensity factors of mode I and mode II, $r$ and $\theta$ are coordinates of the point in vicinity of crack tip.

The stress-optic law for a plane stress case can be expressed as [11]:

$$\tau_{max} = \frac{Nf_\sigma}{2h} \qquad (2)$$

Where $\tau_{max}$ is the maximum shear stress, $N$ is the relative retardation in terms of a complete cycle of retardation (isochromatic fringe order) and $f$ is the material fringe value which was obtained from the calibration test.

The maximum in plane shear stress defines as:

$$(\tau_{max})^2 = (\frac{\sigma_r - \sigma_\theta}{2})^2 + (\tau_{r\theta})^2 \qquad (3)$$

Combining Eq. (2) and Eq. (3) leads to:

$$(\frac{Nf_\sigma}{h})^2 = (\sigma_r - \sigma_\theta)^2 + 4(\tau_{r\theta})^2 \qquad (4)$$

The least square over-determined solving method was employed to calculate the stress intensity factors using the combination of Eq. (1) and Eq. (4) with polar coordinates and fringe numbers ($r$, $\theta$, and $N$) of twenty selected pixels.

### 3.2. FEM analysis

A two-dimensional finite element analysis of the contact problem between the considered half-plane and asymmetrical tilted wedge assuming plane stress condition was performed using ABAQUS FEM package In this model, surface-to-surface contact was employed to simulate the interaction between the wedge and the plate[12]. Some changes like mesh refinement, contact elements instead of regular ones, and application of dynamic load rather than static one were performed to prevent the pressure driven diffusion of the contacting parts. These changes revealed to be effective in eliminating the diffusion at the contact region. Singular elements were also utilized around the crack tip to avoid the infinite stress values. [12] Moreover, the tilted wedge was restricted to move in the vertical direction to prevent application of any external couple. Schematic of the model used in this FEM simulation is shown in Fig. 5.

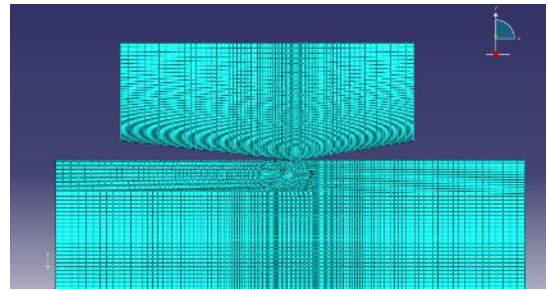

Fig. 5 Schematic of FEM model

## 4. Results and discussion

The intensity factors $K_I$ and $K_{II}$ for the proposed problem were calculated using both photoelastic technique which employed the image processing code for extracting data and finite element method in ABAQUS software. $K_I$ value revealed to be negative in this experience; therefore, $K_I$ has a value of zero and the plate breaks under second fracture mode. The results are shown in Table 3.

Table 3 Result

|  | Exp. Method | FEM | Rel. error |
|---|---|---|---|
| $K_{II}\ (KPa\sqrt{m})$ | 738.8931 | 704.0401 | 4.71% |

The relative difference between the numerical value of $K_{II}$ and the one of experiment is less than 5% which shows that code of image processing is an effective means for reduction of photoelastic experimental results. Since this relative difference is acceptable, the image processing code is validated.

## 5. Conclusion

Using the image processing code in photoelastic stress analysis for finding the stress intensity factors of a contact problem between a half-plane with 45°-edge crack and an asymmetric tilted wedge, was proven that extracting data from captured images using this cod includes fewer errors. The validation of this code was carried out by comparing the intensity factors resulted in experimental method and numerical solution in FEM which showed good agreement with less than 5% relative difference.

## References

[1] Wang Wei-Chung "A digital imaging algorithm for extracting stress intensity factor from the photoelastic fringe pattern" *Engineering Fracture Mechanics*, Volume 36, Issue 5, 1990, Pages 683-696

[2] A.C. Gillies "A co-operative application of expert systems for photoelastic fringe pattern analysis" *Engineering Applications of Artificial Intelligence*, Volume 4, Issue 1, 1991, Pages 35-49

[3] J. Carazo-Alvarez, S.J. Haake, E.A. Patterson "Completely automated photoelastic fringe analysis" *Optics and Lasers in Engineering*, Volume 21, Issue 3, 1994, Pages 133-149

[4] S. Yoneyama, M. Takashi "A new method for photoelastic fringe analysis from a single image using elliptically polarized white light" *Optics and Lasers in Engineering*, Volume 30, Issue 5, November 1998, Pages 441-459

[5] S Yoneyama, M Shimizu, J Gotoh, M Takashi "Photoelastic Analysis with a Single Tricolor Image" *Optics and Lasers in Engineering*, Volume 29, Issue 6, 1 June 1998, Pages 423-435

[6] C.W. Chang, H.S. Lien, J.H. Lin "Determination of reflection photoelasticity fringes analysis with digital image-discrete processing" *Measurement*, Volume 41, Issue 8, October 2008, Pages 862-869

[7] Che-Way Chang, Ping Huang Chen, Hung Sheng Lien "Evaluation of residual stress in pre-stressed concrete material by digital image processing photoelastic coating and hole drilling method" *Measurement*, Volume 42, Issue 4, May 2009, Pages 552-558

[8] T.Y. Chen, P.H. Hou, J.Y. Chiu "Measurement of the ballscrew contact angle by using the photoelastic effect and image processing" *Optics and Lasers in Engineering*, Volume 38, Issues 1-2, July-August 2002, Pages 87-95

[9] M.H. Ghasemi, M. Khaleghian and N. Soltani "Errors Estimation for Evaluating Mixed-Mode Stress Intensity Factors for Cracks Emanating from SharpNotches Using Simulated Photoelasticity" *ISSN* 1818-4952, 2010

[10] D.Broek, "Elementary Engineering Fracture Mechanics" martinus Nijhoff publishers, Columbus.

[11] E. J. Hearn, Photoelasticity, Merrow Publishing Co., Watford, England.

[12] Abaqus 6.10 Online Documentation © Dassault Systèmes, 2010